\documentclass[]{emulateapj}
\usepackage{amsmath}
\usepackage{amssymb}
\usepackage{amstext}
\usepackage{epsfig}

\usepackage{color}
\definecolor{myColor}{rgb}{0.9,0.9,0.9}

\begin{document}

\shorttitle{CMB Aberration}
\shortauthors{Burles and Rappaport}
\title{ABERRATION OF THE COSMIC MICROWAVE BACKGROUND}

\author{S. Burles\altaffilmark{1} and  S. Rappaport \altaffilmark{1}}   

\altaffiltext{1}{Department of Physics and Kavli Institute for Astrophysics and Space Research, MIT, Cambridge, MA 02139; {\tt }}


\begin{abstract}

The motion of the solar system barycenter with respect to the cosmic microwave background (CMB) induces a very large apparent dipole component into the CMB brightness map at the 3 mK level.  In this Letter we discuss another kinematic effect of our motion through the CMB: the small shift in apparent angular positions due to the aberration of light.  The aberration angles are only of order $\beta \simeq 0.001$, but this leads to a potentially measurable compression (expansion) of the spatial scale in the hemisphere toward (away from) our motion through the CMB.  In turn, this will shift the peaks in the acoustic power spectrum of the CMB by a factor of order $1 \pm \beta$.  For current CMB missions, and even those in the foreseeable future, this effect is small, but should be taken into account.  In principle, if the acoustic peak locations were not limited by sampling noise (i.e., the cosmic variance), this effect could be used to determine the cosmic contribution to the dipole term.  

\end{abstract}


\section{Introduction}
\label{sec:intro}

The motion of the solar system barycenter with respect to the cosmic microwave background (CMB) induces a very large apparent dipole component into the CMB brightness map (3.353 mK,
Bennett et al. 1996), and a much fainter, but still readily detectable, quadrupole signature at the $\sim 1\, \mu$K level (e.g., Lineweaver et al. 1996).  The empirically fitted dipole term is subtracted out before computing the spherical harmonics to the residual CMB fluctuations (e.g., Bennett et al. 2003).  The concomitant kinematically induced quadrupole term is considerably smaller than the cosmic quadrupole term and can be subtracted off as well.  One consequence of this operation is that the resultant CMB power spectrum yields no intrinsic dipole term, which would otherwise be of considerable interest for cosmological models.   

The intrinsic cosmic dipole is of particular interest as it represents structure on the largest observable scale, and there may well be interesting phenomena to study on the scales of the lowest multipoles.
Recent analyses of the first year all-sky WMAP datasets confirm the low quadrupole amplitude (Gaztan{\~n}aga et al. 2003;  Bennett et al. 2003; Tegmark et al. 2003)
as was observed in the earlier COBE DMR maps (Smoot et. al. 1992; Hinshaw et al. 1996).
In addition, several papers have pointed out the unlikely alignment of the lowest multipoles for a random Gaussian field, defining the so-called "Axis-of-Evil" (Land \& Magueijo 2005; de Oliveira-Costa et al. 2004; Copi et al. 2004; Bielewicz et al. 2005).   Possible explanations have been suggested, from selection of foreground-free sky (Slosar \& Seljak 2004) to weak lensing of the cosmic dipole by local structures (Vale 2005).  Future CMB datasets may settle the matter, but a measurement or constraint of the amplitude and direction of the intrinsic cosmic temperature dipole could be essential to reach a final resolution.

In this Letter we discuss another kinematic effect of our motion through the CMB, i.e., the small shift in apparent angular positions due to the aberration of light (Bradley 1729).  The aberration angles are only of order $ \beta$ (where $\beta = 0.00123$; Lineweaver et al. 1996), but this leads to a potentially measurable compression (expansion) of the spatial scale in the hemisphere toward (away from) our motion through the CMB.  In turn, this will shift the peaks in the angular temperature power spectrum of the CMB by a factor of order $1 \pm \beta$.  The effect amounts to about 1 multipole ($\ell$) bin at the 5th harmonic of the CMB acoustic spectrum.  In any event, the aberration effect can and should be corrected even though it leads to shifts in the position of the CMB features by only up to $\sim 4^\prime$.  If it were not for inherent uncertainties in determining the centroids of the acoustic peaks due to sampling limitations (i.e., the cosmic variance), this effect could, in principle, yield an independent measurement of the kinematic contribution to the dipole and allow a correction to measure the intrinsic power in the cosmic dipole.

\section{The Aberration Effect on the CMB}
\label{sec:aber}

We can describe the aberration of the CMB radiation as follows.  If we use spherical coordinates and define $\theta$ as the polar angle with respect to the direction of motion, and $\phi$ as the azimuthal angle, then the transformation of angles from the CMB to the barycenter frame is:
\begin{eqnarray}
\sin\theta & = & \frac{\sin\theta^\prime}{\gamma(1-\beta \cos\theta^\prime)} ~~~,\\
\phi & = & \phi^\prime ~~~,
\end{eqnarray}
where the unprimed and primed frames are the CMB and barycenter frames, respectively, $\beta$ is the velocity of the barycenter with respect to the CMB ($\sim$0.001), $\gamma$ is the usual Lorentz factor, and where $\theta=0$ corresponds to the `forward' direction.  For small $\beta$ the $\theta$ transformations can be expanded in a Taylor series to lowest order in $\beta$ as:
\begin{eqnarray}
\sin\theta = \sin\theta^\prime (1 + \beta \cos \theta^\prime) ~~~.
\end{eqnarray}
Finally, we can expand the arc sine function to find $\theta(\theta^\prime)$:
\begin{eqnarray}
\theta  =  \theta^\prime + \beta \sin \theta^\prime ~~~.
\end{eqnarray}
At any location on the celestial sphere, this leads to a stretching or compression of the scales in the $\theta$ and $\phi$ directions by fractional amounts:
\begin{eqnarray}
\frac{d\theta}{d\theta^\prime} & = & 1+\beta \cos \theta^\prime \\
\frac{\sin \theta d \phi}{\sin \theta^\prime d \phi^\prime} & = & 1 + \beta \cos \theta^\prime ~~~.
\end{eqnarray}
From this we conclude that the two-dimensional angular scale size is compressed by a factor of 
$1 - \beta \cos \theta^\prime$ in the forward hemisphere, and stretched by a factor of $1 + \beta |\cos \theta^\prime|$ in the backward direction.  
 
\section{Shift in Location of the CMB Acoustic Peaks}
\label{sec:loc}

The full-sky map of the temperature anisotropies as measured by the WMAP satellite (Bennett et al. 2003; Page et al. 2003) is well described by adiabatic fluctuations in a $\Lambda$CDM cosmology (Spergel et al. 2003) after removal of foreground contamination to the cosmic signal (Hinshaw et al. 2003; Tegmark et al. 2003).  Using the current concordance model of $\Lambda$CDM with the Boltzmann code CMBFAST (Seljak \& Zaldarriaga 1996), one can accurately predict the expected angular power spectrum of primary temperature anisotropies that will be measured by the Planck satellite  (Balbi et al. 2002; Piat et al. 2002; Sandri 2004) out to multipoles of $\sim$1500.  For the discussion below, we will assume this spectrum for the cosmic angular power spectrum in the CMB.  

If the CMB power spectra are computed separately for portions of the forward and backward looking regions where the angular scale stretching factors are approximately constant, e.g., for $ \theta^\prime < 45^\circ$ and $\theta^\prime > 135^\circ$, then the $\ell$ values of the spherical harmonics at the acoustic peaks should shift (in opposite directions) for the segments of the celestial sphere located in the forward and backward directions.  The shift in $\ell$ values is expected to be $\delta \ell_{f,b} = \pm \overline{\beta} \ell$, where $\overline{\beta}$ represents the average value of $\beta |\cos \theta^\prime|$ over the regions of interest ($\overline{\beta} \simeq 0.001$).
Thus, in order to detect this effect -- at all -- requires an accuracy in determining the centroids of the peaks in the CMB harmonic structure to about one part in 1000.  

The uncertainty in measuring the amplitude $C_\ell$ due to the finite number of modes on the celestial sphere is given by:
\begin{eqnarray}
\frac{\delta C_\ell}{C_\ell} = \sqrt{\frac{2}{(2\ell + 1)f_{\rm sky}}} ~~ \simeq ~~ \sqrt{\frac{1}{\ell f_{\rm sky}}} ~~~,
\end{eqnarray}
(Scott et al. 1994) where $f_{\rm sky}$ is the fraction of the celestial sphere included in the analysis, and where the experimental noise contribution is assumed to be negligible compared to the cosmic variance.  In the nth harmonic peak, centered at $\ell = \ell_n$ there are $\sim \ell_n/n$ values of $C_\ell$, all of which can be determined with roughly comparable fractional accuracy --- given by eq. (7).  This implies that the peak's centroid can be determined in an ideal measurement with an uncertainty of approximately:
\begin{eqnarray}
\delta \ell  \simeq  \left(\frac{\ell_n}{n}\right) \frac{1}{\sqrt{\ell_n f_{\rm sky}}} \frac{1}{\sqrt{\ell_n/n}} \simeq \frac{1}{\sqrt{n f_{\rm sky}}} ~~~,
\end{eqnarray} 
For observational hemispherical caps in the forward and backward directions of $\theta^\prime \sim 45^\circ$, the uncertainty in $\ell_n$ is about $\sqrt{1/(5f_{\rm sky})} \simeq 1$ for the 5th harmonic. And, since the expected shift in $\ell$ due to the aberration effect is of order $\delta \ell \simeq {\overline \beta} \ell$, this implies that by $\ell \sim 1000$ we may expect to detect a significant shift in $\ell_n$ between the forward and backward hemispheres.  If we utilize all the peaks up to the 5th harmonic, then we gain a small additional factor in signal to noise.  A more formal error estimate is discussed below.

\begin{figure}[t]
\begin{center}
\epsfig{figure=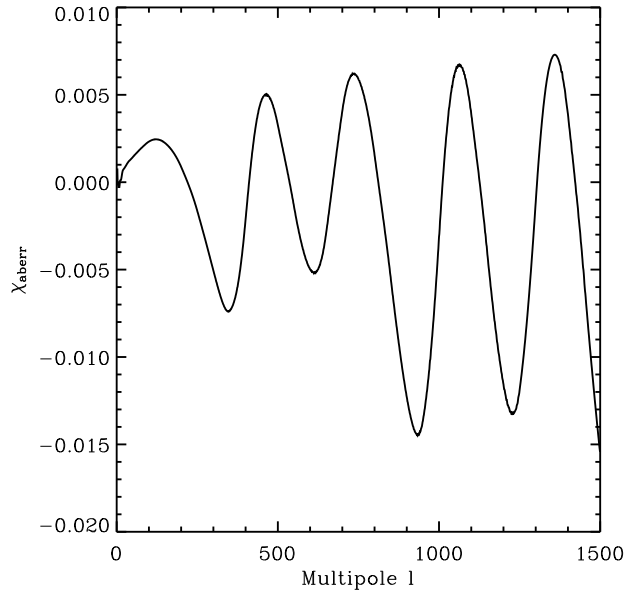,width=0.47\textwidth}
\caption{Theoretical plot of $\chi_{\rm aberr}$ given by equation (9). Typical error bars on each of the $\sim$1500 values of $C_\ell$ used to construct such a plot from actual data would be given by eq. (7) and are of order $\pm 10\%$ in the ratio of $C_\ell^f/C_\ell^b$, or about $\pm 0.1$ in the $\ln$.  These uncertainties about 10 times larger than the aberration effect on individual values of $\chi_{\rm aberr}$, but the overall signature of the shifts should be detectable with all 1500 points (see text).  
}
\end{center}
\end{figure}

Figure 1 illustrates how the aberration effect would quantitatively affect the CMB power spectrum peaks.  What is shown is a plot of the quantity
\begin{eqnarray}
\chi_{\rm aberr} & = \ln \left[\displaystyle{\frac{C_\ell^f \ell_f (\ell_f + 1)}{C_\ell^b \ell_b (\ell_b+1)}} \right] ~~~, \\
                             & \simeq 2 \overline{\beta}  \left[ \displaystyle \frac{ d ln C_\ell}{d ln \ell} + 2 \right ] ~~~, 
\end{eqnarray} 
which is simply a convenient way of visualizing the effects of the aberration shifts.  A $\Lambda$CDM model was assumed in Figure 1, and was obtained through the package of CMBEASY (Doran 2005), but the generic nature of the derivative of the angular power spectrum in the presence of adiabatic fluctuations would be very similar for other models.  We have carried out a formal statistical analysis of the uncertainties in determining $\beta$ using the function given by eq. (9) (see Fig. 1) and the assumed uncertainties in the individual $C_\ell$'s given by eq. (7).  We assumed that $C_\ell^f$ and $C_\ell^b$ were both determined over 45$^\circ$ cones on the celestial sphere in the forward and backward directions, respectively.  The result is that the formal $1\sigma$ uncertainty in $\overline{\beta}$ is $3 \times 10^{-4}$, and thus the aberration effect should be detectable at a confidence level above 99.9\%.  A direct estimation  of the significance of the ratio in Fig. 1 gives a $2\sigma$ detection out to $\ell=1000$, and a $3.3\sigma$ detection including all multipoles out to $\ell = 1500$.  This is a slight overestimate as we have only included noise due to cosmic variance, but an analysis of the full-sky would provide a measurement at slightly higher significance.

There are two competing effects in trying to choose an optimum fractional hemispherical cap size in the forward and backward directions in order to search for the kinematic aberration effect.  In the first, the larger the solid of angle of the cap size (as characterized by $\theta^\prime_{\rm max}$) the greater is the available statistical precision.  The fractional solid angle in either of the hemispheres goes as $f _{\rm sky}= (1-\cos \theta^\prime)/2$.  On the other hand, the larger $\theta^\prime_{\rm max}$ is, the smaller the average value of the shifts in the centroids of the acoustic peaks.  Values of $\theta^\prime_{\rm max}$ of $30^\circ$ and $45^\circ$, for example, yield 6.7\% and $14.6\%$ of the celestial sphere, respectively, while preserving 93\% and 85\% of the aberation effect, when averaged over the respective hemispherical caps.

\section{Summary and Conclusions}
\label{sec:sum}

We have shown that the aberration of the detected CMB radiation, due to the motion of the barycenter relative to the CMB, can shift the centroids of the acoustic peaks in the forward vs. the backward directions by a measurable amount.  This effect may already be marginally detectable with the WMAP data, and should be detectable with the data to be acquired with the Planck Mission.  To put the magnitude of this aberration effect into perspective, it is anticipated that the Planck CMB maps will be constructed with $\sim 10^\prime$ angular bins (Sandri et al. 2004).  Locations in CMB features will be shifted by $2.^\prime 1$, $3.^\prime 0$, and $4.^\prime 2$ at $\theta^\prime = 30^\circ$, $45^\circ$, and $90^\circ$, respectively.

If there is, in fact, a significant measurable difference between the locations of the peaks in the CMB power spectrum for the forward and backward directions, then this will serve to demonstrate the expected kinematic effect due to motion of the barycenter.  At present, it does not appear possible to measure this shift with better than about 1 part in 4000 accuracy due to the cosmic variance.  It may be worth noting, nonetheless, that if it {\em were} somehow possible to achieve a factor of $\sim$100 better in accuracy, this would be sufficient to allow the kinematic contribution to the dipole term to be determined independently with $\sim$1\% accuracy.  This, in turn would enable a determination of the cosmic contribution to the dipole term.  

Also, we would like to emphasize that a detection and measurement of the aberration of the CMB is complementary to other effects which arise due to the motion of the solar system. In particular, the intensity quadrupole explored by Kamionkowski \& Knox (2003) presents an independent observable to constrain the contribution of the kinematic temperature dipole.

Finally, we remark that in the same spirit of searching for direction dependent aberration effects,
it might be generally worthwhile to measure the CMB power spectrum over a number ($\sim$10) of independent regions of the sky and intercompare the results.  The isotropy of the temperature power spectrum has been investigated (Eriksen et al. 2003, Hansen et al. 2004), and regions of excess asymmetry have been reported.  The next generation of isotropic tests should account for the aberration of the CMB to avoid systematic errors over the sky.   

\acknowledgements 

We thank Max Tegmark and Scott Hughes for helpful discussions.
SR acknowledges support from NASA Chandra Grant NAG5-TM5-6003X.
SB acknowledges support from NSF Grant AST-0307705.

\vspace{1 cm}

\end{document}